\documentclass[aps,prd,reprint,nofootinbib]{revtex4-2}

\usepackage{amsmath,amssymb,bm}
\usepackage{graphicx}
\usepackage{xcolor}
\usepackage[colorlinks=true,linkcolor=blue,citecolor=blue,urlcolor=blue]{hyperref}
\usepackage{soul}
\usepackage{orcidlink}  

\begin{document}

\title{
Sensitivity of Heavy-Quark Dipolar Flow to its Initial Spatial Distributions in Cu+Au Collisions}


\author{Ankit Kumar Panda \orcidlink{0000-0002-9394-6094}}
\email{ankitkumarpanda932@gmail.com}
\affiliation{Key Laboratory of Quark and Lepton Physics (MOE) \& Institute of Particle Physics, Central China Normal University, Wuhan 430079, China}

\author{Tribhuban Parida \orcidlink{0000-0002-9419-2247}}
\email{tribhu.451@gmail.com}
\affiliation{AGH University of Krakow, Faculty of Physics and Applied Computer Science, aleja Mickiewicza 30, 30-059 Cracow, Poland}

\date{\today}


\begin{abstract}
We investigate charm-quark dynamics in asymmetric Cu+Au collisions at top RHIC energy using a Langevin approach embedded in a realistic hydrodynamic background. The intrinsic asymmetry of the colliding nuclei leads to a spatially lopsided initial energy-density profile, which generates a dipolar flow structure in the transverse plane even at midrapidity. As charm quarks propagate through this medium, they acquire a finite directed flow, $v_1$.
We find that the $p_T$-integrated heavy-quark $v_1$ is approximately an order of magnitude larger than that of charged hadrons. In addition, the $p_T$-differential $v_1$ exhibits strong sensitivity to the initial spatial distribution of heavy quarks, emphasizing the importance of pre-equilibrium dynamics in determining final-state anisotropies. Beyond geometric effects, $v_1$ also provides direct sensitivity to medium interactions through the temperature-dependent drag coefficient. Its pronounced dependence on this transport input indicates that precision measurements of heavy-flavor directed flow could place meaningful constraints on heavy-quark transport coefficients, thereby improving Langevin-based descriptions and predictive power for heavy-flavor observables in heavy-ion collisions.
\end{abstract}

\maketitle

\section{Introduction}{\label{intro}}

Heavy quarks serve as powerful probes of the quark--gluon plasma (QGP)~\cite{Song:2010mg,Busza:2018rrf,Kolb:2000sd,Heinz:2013th} produced in high-energy heavy-ion collisions. Created predominantly in initial hard scatterings, charm and bottom quarks traverse the entire space-time evolution of the medium before hadronization~\cite{Rapp:2008tf,He:2014cla,Andronic:2021erx}. Their large masses ensure early production and sustained interaction with the expanding plasma, making heavy-flavor observables particularly sensitive to the transport properties and dynamical evolution of the QGP~\cite{Rapp:2018qla,Cao:2013ita,Cao:2018ews,Scardina:2017ipo,Das:2009vy}. This sensitivity is reflected in measurements of the nuclear modification factor ($R_{AA}$) and anisotropic flow coefficients~\cite{Moore:2004tg,Olsen:2017bmm,Harris:2023tti,STAR:2018zdy,Das:2015ana,Alberico:2013bza,Cao:2015hia,He:2011qa}.

Among these observables, the directed flow coefficient
$v_1(y_p,p_T)=\langle \cos(\phi-\Psi)\rangle$, 
quantifies the first harmonic of the azimuthal particle distribution, where $\phi$ is the particle azimuthal angle, $y_p$ and $p_T$ denote rapidity and transverse momentum, and $\Psi$ is the reference-plane angle. As an early-time observable, $v_1$ is sensitive to a variety of mechanisms, including the three-dimensional initial energy-density profile, event-by-event geometric asymmetries, and electromagnetic fields generated in non-central collisions~\cite{NA49:2003njx,Shen:2020jwv,Gursoy:2014aka,STAR:2014clz,Panda:2023akn}. In particular, heavy-flavor $v_1$~\cite{Das:2016cwd,Sun:2023adv,Chatterjee:2018lsx,Panda:2026kko} provides enhanced sensitivity to the early-time propagation of heavy quarks through the evolving QGP background.

In non-central collisions, the bulk medium develops a forward--backward asymmetric tilted geometry due to unequal energy deposition in the overlap region. Heavy quarks, however, are produced in initial binary nucleon--nucleon collisions and approximately follow the binary-collision profile, leading to a spatial mismatch between the heavy-quark distribution and the tilted medium at finite rapidity. Within Langevin dynamics coupled to hydrodynamic evolution, this mismatch generates a finite heavy-quark directed flow. In this framework, heavy-flavor $v_1$ typically inherits the sign of the bulk flow while attaining a substantially larger magnitude~\cite{Chatterjee:2017ahy}. 

Since heavy quarks are produced before local thermalization, they also undergo a pre-equilibrium stage prior to hydrodynamic evolution, during which diffusion and momentum broadening can modify their phase-space distribution. Because directed flow originates from the relative spatial configuration between heavy quarks and the asymmetric medium, such early-time modifications can significantly influence the final $v_1$, making it sensitive to both the initial spatial distribution and pre-equilibrium dynamics. Motivated by this sensitivity, we systematically investigate the influence of different heavy-quark initialization schemes on $v_1$ and explore whether future measurements can constrain possible early-time modifications. While pre-equilibrium effects on heavy-flavor observables such as $R_{AA}$ and $v_2$ have been studied previously~\cite{Das:2017dsh,Singh:2025duj,Das:2015aga,Sun:2019fud,Ruggieri:2013bda}, their impact on directed flow remains largely unexplored.

Directed flow can arise from multiple sources, including tilted initial geometries \cite{Bozek:2010bi}, initial-state fluctuations \cite{Luzum:2010fb}, and possible charge-dependent deflections induced by strong electromagnetic fields in non-central collisions~\cite{Gursoy:2014aka,STAR:2016cio,STAR:2023jdd,Jia:2014jca,Panda:2023akn,Zhang:2025tgt}. Phenomenological tilted-fireball models~\cite{Bozek:2010bi,Parida:2022lmt} typically introduce the required asymmetry through additional parameters, leading to a degree of model dependence even when constrained by charged-hadron directed flow \cite{Chatterjee:2018lsx, Chatterjee:2017ahy}. By comparison, asymmetric collision systems such as Cu+Au~\cite{Voronyuk:2014rna,PHENIX:2018hho,Nakamura:2022idq} naturally generate an uneven transverse geometry due to the different sizes and density profiles of the colliding nuclei \cite{Bozek:2012hy,PHENIX:2015zbc}, providing a more constrained and physically transparent framework for studying heavy-flavor directed flow. The intrinsic dipolar deformation of the initial energy density in such systems generates a sizable rapidity-even $v_1$~\cite{Teaney:2010vd,Luzum:2010fb,Gardim:2011qn,Bozek:2012hy,Shafi:2025feq} near midrapidity, where experimental measurements benefit from larger particle yields and improved statistical precision. The present work focuses on this geometrically driven rapidity-even component. Although event-by-event fluctuations are not included in our framework, a finite $v_1$ still emerges at midrapidity purely from the underlying geometric asymmetry, implying that the observed signal is dominated by the collective response to the asymmetric initial geometry. By contrast, the rapidity-odd component originates from forward--backward asymmetry in particle production, changes sign across midrapidity, and vanishes at $y_p \simeq 0$.

This paper is organized as follows. In Section~\eqref{dynamicsandinit}, we describe the heavy-quark dynamics in the hydrodynamic background, including the initial conditions, collision geometry, and medium evolution setup. Section~\eqref{results} presents the numerical results and discusses the effects of different spatial sampling schemes on heavy-quark observables. Finally, Section~\eqref{conclusion} summarizes the main findings and outlines possible directions for future investigations.
\section{Heavy-Quark Dynamics in a Hydrodynamic Medium}{\label{dynamicsandinit}}

We investigate the evolution of heavy quarks, particularly charm quarks, in the hot and dense medium created in asymmetric Cu+Au collisions at top RHIC energy by combining relativistic hydrodynamics with Langevin dynamics. The hydrodynamic evolution provides the spacetime profiles of the fluid velocity, $u^\mu(x)$, and temperature, $T(x)$, which together define the dynamical background through which heavy quarks propagate. The medium evolution is initialized using energy density profiles obtained from a Monte Carlo Glauber model with 20,000 events, ensuring a smooth initial condition, with the hydrodynamic evolution starting at an initialization time of $\tau_0 = 0.4$~fm. In our setup, the Cu nucleus is positioned at $-b/2$ and the Au nucleus at $+b/2$ along the impact parameter axis ($x$-axis). We emphasize that this choice of orientation is purely conventional and does not affect the physics or the conclusions presented in this work. The shear viscosity to entropy density ratio is fixed at $\eta/s = 0.08$, while the bulk viscosity $\zeta/s$ is taken to be temperature dependent~\cite{Moreland:2018gsh}. The hydrodynamic evolution is continued until the local temperature reaches 145 MeV, below which hadronization is assumed to take place.

The propagation of heavy quarks through this evolving medium is described using Langevin dynamics~\cite{He:2022ywp,vanHees:2005wb,Cao:2011et,Scardina:2017ipo}, which, in the post-point realization, can be written as
\begin{align}\nonumber
\Delta \mathbf{r}_i &= \frac{\mathbf{p}_i}{E} \Delta t, \label{eq:Langevin_x} \\ 
\Delta \mathbf{p}_i &= -\gamma(T) \mathbf{p}_i \Delta t + \rho_i \sqrt{2 D \Delta t}, 
\end{align}
where $\Delta \mathbf{r}_i$ and $\Delta \mathbf{p}_i$ denote the changes in the position and momentum of a heavy quark during a time step $\Delta t$, with $i=x,y,z$. Here, $\gamma(T)$ and $D$ represent the drag and diffusion coefficients, respectively, which characterize the interaction of heavy quarks with the medium. The drag coefficient is taken to be temperature dependent in order to reflect the evolving properties of the QGP, and we explicitly investigate the sensitivity of our results to different temperature dependencies of $\gamma(T)$. The stochastic term $\rho_i$ is sampled independently from a Gaussian distribution at each time step and satisfies $\langle \rho_i \rangle = 0$ and $\langle \rho_i \rho_j \rangle = \delta_{ij}$. The drag and diffusion coefficients are connected through the fluctuation-dissipation theorem, and for a thermalized medium the diffusion coefficient is chosen as $D = \gamma(T) E T$, where $E = \sqrt{\mathbf{p}^2 + m^2}$ is the heavy-quark energy. This choice ensures that the heavy-quark phase-space distribution approaches the equilibrium Boltzmann-Juttner distribution in the long-time limit~\cite{He:2013zua}.
Since the Langevin evolution is performed in the local rest frame of the fluid, the heavy-quark phase-space coordinates are Lorentz transformed to the local fluid frame at each time step and subsequently transformed back to the laboratory frame after the momentum update.

To explore the sensitivity of heavy-quark evolution to the initial conditions, we consider three different sampling schemes for the initial phase-space distribution: (i) sampling according to the energy density profile of the medium, (ii) sampling based on the binary collision ($N_{\rm coll}$) distribution associated with the initial hard scatterings, and (iii) sampling from a uniform box-type distribution within the region $x,y \in [-2,2]$~fm, the details of which will be discussed in the next section with a pictorial depiction showing a sampling comparison in Fig.\eqref{fig:normalised}. The initial momentum distribution of heavy quarks is generated using the Fixed Order plus Next-to-Leading Logarithm (FONLL) framework~\cite{Cacciari:2005rk,Cacciari:2012ny}, which successfully reproduces the $\mathrm{D}$-meson spectra measured in proton--proton collisions after fragmentation~\cite{Scardina:2017ipo}. In total, $5 \times 10^6$ heavy quarks are sampled to ensure sufficient statistical precision in the analysis.
To quantify the collective motion of heavy quarks, we compute the event-averaged directed flow, $v_1 = \langle p_x/p_T \rangle$. The results presented in Sec.~\eqref{results} illustrate the effects of different initial sampling schemes, impact parameters, and temperature dependencies of the drag coefficient on the heavy-quark distributions and directed flow. In Fig.~\eqref{fig:energydensity}, we show the initial energy density profile at $b = 6$~fm, which serves as the input for the subsequent hydrodynamic evolution performed using the MUSIC-Hydro framework~\cite{Schenke:2010nt,Schenke:2010rr,Paquet:2015lta}. Although additional calculations are carried out for other impact parameters to study the influence of varying initial geometries, they are not shown here. Comparisons between different sampling prescriptions, momentum spectra, and drag-coefficient parametrizations provide insight into the interplay between the initial geometry, medium evolution, and drag-diffusion dynamics in shaping heavy-quark collectivity.

\section{Results and discussions}{\label{results}}

\begin{figure}[h]
    \centering
    \includegraphics[width=\linewidth]{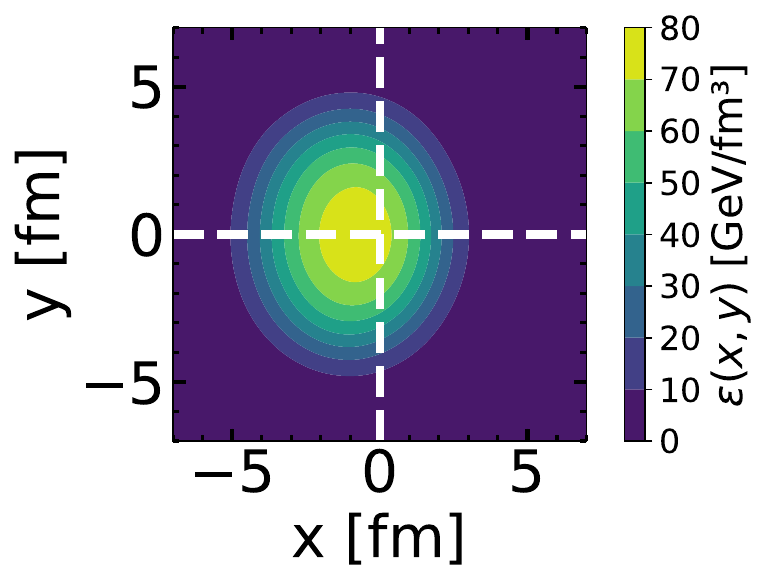}\\
    \caption{Initial energy density profile for a Cu+Au collision at $b = 6$~fm, with Cu at $-b/2$ and Au at $b/2$, used as input for MUSIC hydrodynamic evolution at $\sqrt{s_{NN}} = 200$~GeV.}
    \label{fig:energydensity}
\end{figure}

\begin{figure}[h]
    \centering
    \includegraphics[width=\linewidth]{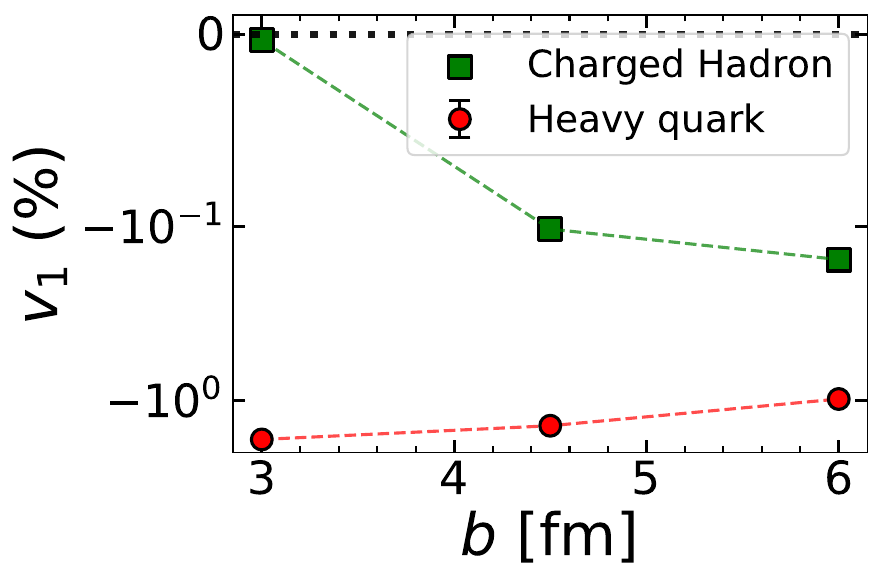}
    \caption{Comparison of integrated $v_1$ versus impact parameter for light ($p_T \in [0,3]$~GeV) and heavy ($p_T \in [0,10]$~GeV) quarks, calculated with $\gamma (T) = 0.6\,T$ using $N_{\rm coll}$ sampling.}
    \label{fig:v1integrated}
\end{figure}

The initial energy density profile exhibits a pronounced asymmetry in the transverse plane, reflecting the intrinsic size difference between the colliding nuclei. In the collision geometry, the Cu nucleus is shifted by $-b/2$ and the Au nucleus by $+b/2$ along the impact parameter direction ($x$-axis), as discussed earlier. Due to its smaller transverse thickness, the Cu nucleus deposits less energy compared to the thicker Au nucleus, resulting in a spatially lopsided energy density distribution, as illustrated in Fig.~\eqref{fig:energydensity}.
This asymmetry is most prominent along the impact parameter direction: the energy density varies significantly along the $x$-axis, while remaining approximately symmetric along the $y$-axis. As a consequence, the pressure gradients generated during the subsequent hydrodynamic evolution are themselves asymmetric along the positive and negative $x$-directions.  
Such an initial asymmetry drives a preferential collective expansion from the negative toward the positive $x$-direction, leading to an imbalance in the final-state transverse momentum component $p_x$. This momentum anisotropy gives rise to a non-zero directed flow ($v_1$) of charged hadrons. The $p_T$-integrated charged-hadron $v_1$ for different impact parameters is shown as square markers in Fig.~\eqref{fig:v1integrated}.  
As collisions become more peripheral, the geometric asymmetry of the overlap region becomes more pronounced, amplifying the asymmetry in the pressure gradients and, consequently, the directed flow signal. Accordingly, the magnitude of charged-hadron $v_1$ increases from central to peripheral collisions, as illustrated in Fig.~\eqref{fig:v1integrated}.  

It should be noted that the directed flow presented here is computed with respect to the reaction plane (see Sec.~\eqref{intro}). In our simulations, the reaction plane angle is set to zero by aligning the impact parameter vector along the $x$-axis. Experimentally, the reaction plane angle—defined as the angle between the reaction plane and the laboratory $x$-axis—is not directly measurable, but can be inferred from the spectator plane reconstructed using spectator nucleons \cite{STAR:2011hyh}.

\begin{figure}[h]
    \centering
    \includegraphics[width=\linewidth]{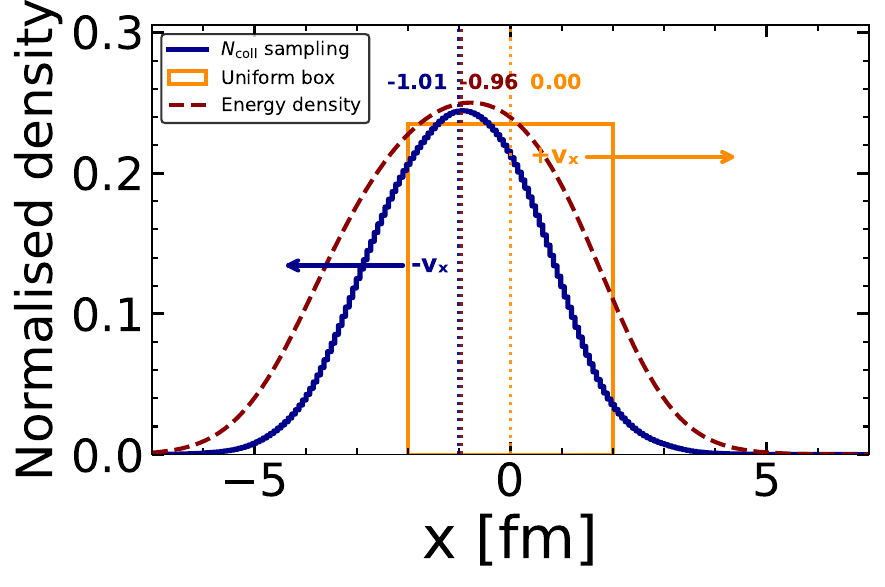}
    \caption{Initial heavy-quark distributions along the impact-parameter ($x$) axis at $y = 0$ for a Cu+Au collision with $b = 6$~fm at $\sqrt{s_{\rm NN}} = 200$~GeV. Shown are three distributions: two from the Monte Carlo Glauber model and one uniform (box-type) distribution, with their means $-1.01$, $-0.96$, and $0.00$ for $N_{\rm coll}$, energy density, and uniform box samplings respectively.}
    \label{fig:normalised}
\end{figure}

To study heavy-quark dynamics in such an asymmetric medium, we initialize their production points according to the binary collision ($N_{\rm coll}$) distribution and evolve them using Langevin dynamics on the hydrodynamic background. The schematic sampling using the real data is shown in Fig.~\eqref{fig:normalised}, where we display the energy density distribution along the $x$-axis at $y=0$, together with the corresponding binary collision distribution projected onto the same axis.
A clear difference emerges when comparing the spatial distributions of the medium energy density, which drives the flow of charged hadrons, with the heavy-quark production points obtained from $N_{\rm coll}$. This offset between the mean positions of the medium and heavy-quark distributions plays a decisive role in shaping the final directed flow. 

Heavy quarks originating from different locations within the medium experience unequal interactions, undergoing asymmetric drag along the positive and negative $x$-directions, which gives rise to a finite heavy-quark $v_1$.  
The resulting $p_T$-integrated charm-quark $v_1$, calculated within our framework, is shown alongside that of charged hadrons in Fig.~\eqref{fig:v1integrated}. When compared to the light-flavor dominated charged hadrons, the heavy-quark $v_1$ is roughly an order of magnitude larger while retaining the same sign across impact parameters up to $b=6$~fm. This behavior can be traced back to the relative spatial distributions illustrated in Fig.~\eqref{fig:normalised}: the heavy-quark production profile is shifted to the left of the energy density distribution (mean $-1.01$ versus $-0.96$ for the energy density), indicating that a larger fraction of heavy quarks initially resides on one side of the medium and consequently experiences asymmetric drag during their evolution. This effect is quantified through the mean values indicated in Fig.~\eqref{fig:normalised}. Furthermore, the difference in their mean in these distributions gradually increases toward more central collisions, amplifying the difference in the most central events.

\begin{figure}[h] 
\centering \includegraphics[width=\linewidth]{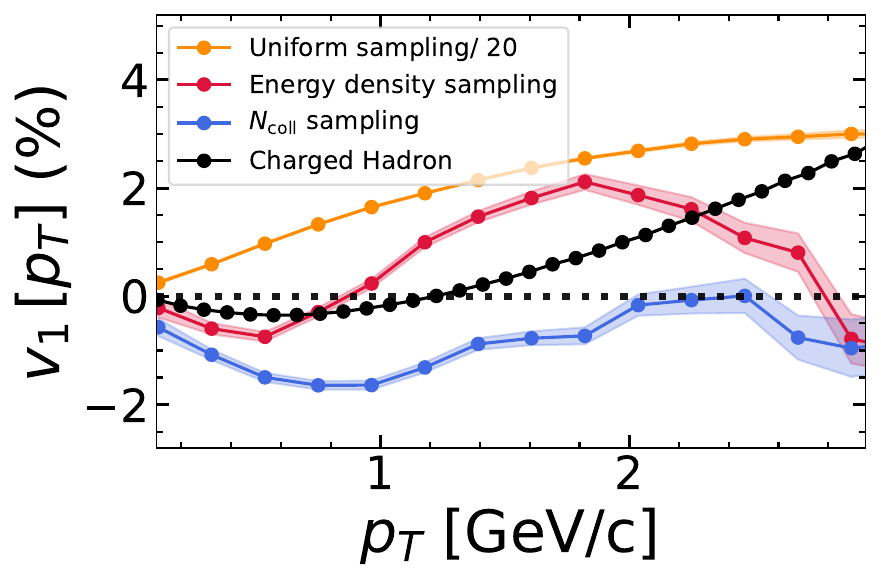} \caption{Comparison of $v_1$ for different heavy-quark sampling methods at $b = 6$~fm, $y_p$ $\in$ [-1,1] with $\gamma (T)= 1.2\,T$. Shaded bands indicate statistical uncertainties.} \label{fig:v1evensample} \end{figure}

Motivated by the early production of heavy quarks and the observed sensitivity of their directed flow ($v_1$) to the relative spatial distributions of heavy quarks and the medium energy density, we investigate how different initialization schemes influence final-state heavy-flavor observables. For this we consider three initialization procedures. In the first scenario, heavy quarks are sampled according to the binary collision ($N_{\rm coll}$) distribution, representing the conventional assumption based on production through initial hard scatterings. In the second scenario, heavy quarks are initialized following the energy density profile of the medium. Finally, in the third scenario, heavy quarks are uniformly sampled within a box centered at $(x,y)=(0,0)$ in the transverse plane.  
These three distributions are illustrated in Fig.~\eqref{fig:normalised}, providing a clear comparison of the relative positions of heavy quarks with respect to the medium and offering intuition for the expected sign and magnitude of the resulting directed flow.  
The corresponding heavy-quark $v_1(p_T)$ for each initialization scheme is presented in Fig.~\eqref{fig:v1evensample} for collisions with impact parameter $b=6$~fm and a temperature-dependent drag coefficient $\gamma (T)= 1.2\,T$. For reference, the $v_1(p_T)$ of charged hadrons obtained from the MUSIC hydrodynamics simulation is also shown. 

The results indicate that the heavy-quark directed flow exhibits a strong dependence on the initialization procedure, displaying qualitatively different behaviors across the scenarios. When heavy quarks are sampled uniformly in the box, most are initially located to the right of the energy density maximum, resulting in a strongly positive $v_1$. In contrast, $N_{\rm coll}$ sampling positions the majority of heavy quarks to the left of the maximum, yielding a negative $v_1$. Sampling along the energy density profile places heavy quarks near the medium peak, producing a $v_1$ that transitions from negative at low $p_T$ to positive at intermediate $p_T$ a similar trend shown by the light charged hadron's $v_1$.  

These findings demonstrate that both the sign and magnitude of $v_1$, as well as its transverse momentum dependence, are strongly influenced by the initial spatial distribution of heavy quarks. Measuring the directed flow ($v_1$) of heavy quarks in asymmetric collisions can give us valuable clues about where heavy quarks are initially located in the collision zone. Even though heavy quarks are created very early in hard scatterings, we start tracking them only when the hydrodynamic evolution of the medium begins. Between their creation and this starting time, there is a pre-equilibrium stage during which the heavy quarks’ positions may change, which can affect the final $v_1$ seen in experiments. By studying heavy-flavor $v_1$, future experiments could tell us whether the actual distribution of heavy quarks at the start of hydrodynamics matches the expected $N_{\rm coll}$ pattern or if it has been modified during the pre-equilibrium phase.

\begin{figure}[h]
    \centering
    \includegraphics[width=\linewidth]{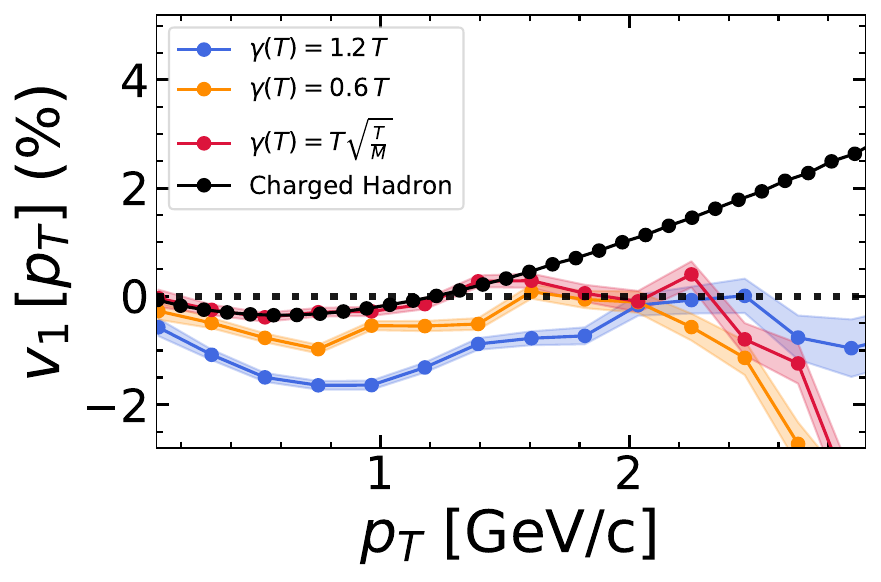}
    \caption{Comparison of $v_1$ for heavy quarks and light charged hadrons at $b = 6$~fm, $y_p$ $\in$ [-1,1] showing sensitivity to the temperature dependence of the drag coefficient. Heavy quarks are sampled via $N_{\rm coll}$; shaded bands indicate statistical uncertainties.}
    \label{fig:v1evenT}
\end{figure}

The directed flow of heavy quarks is also influenced by their interactions with the medium, which are described by the Langevin drag coefficient. By exploring different parameterizations of the drag, we can investigate how $v_1$ reflects both the initial geometry and heavy-quark transport properties. In this study, the drag coefficient is defined as
$\gamma (T)= \gamma_0 T \left(\frac{T}{m}\right)^x$,
where $\gamma_0$ sets the overall interaction strength (which takes values like 0.6 and 1.2 in our current study), and $x$ determines its modified temperature dependence, taking values $0$ or $1/2$. Using these variations, we calculate $v_1$ while keeping the initial heavy-quark distribution fixed to $N_{\rm coll}$ sampling.  
The resulting heavy-flavor $v_1$ is shown in Fig.~\eqref{fig:v1evenT}. While the overall trend of $v_1$ remains similar, its magnitude increases with stronger drag, highlighting the sensitivity of heavy-quark directed flow to the transport properties of the medium.  

The drag coefficient affects not only heavy-quark directed flow but also other key observables, such as the nuclear modification factor $R_{AA}$ and the elliptic flow $v_2$ \cite{Chatterjee:2017ahy}. Among these, the directed flow $v_1$ offers an independent and complementary probe of heavy-quark transport. Unlike $R_{AA}$ and $v_2$, which mainly reflect energy loss and momentum anisotropy, $v_1$ is directly sensitive to the relative spatial configuration of heavy-quark production and the asymmetric medium. Its ability to change sign makes it particularly responsive to both the initial geometry and the transport dynamics. A combined analysis of $v_1$, $R_{AA}$, and $v_2$ can therefore reduce uncertainties in the temperature dependence of heavy-quark transport and improve the overall modeling of heavy-quark evolution in the medium.

We emphasize that the present results are not directly for $D$ mesons, as hadronization via fragmentation or coalescence, as well as subsequent hadronic rescattering and freeze-out effects, are not included in this study; while these effects may modify the results quantitatively, the qualitative features and underlying trends are expected to remain unchanged.


\section{Conclusion}{\label{conclusion}}

We have investigated the directed flow ($v_1$) of heavy quarks in asymmetric Cu+Au collisions at $\sqrt{s_{\rm NN}} = 200$~GeV using hydrodynamics coupled with Langevin dynamics. Our results demonstrate that heavy-quark $v_1$ is highly sensitive to their initial spatial distribution relative to the medium energy density. Depending on the initialization scheme—binary collision ($N_{\rm coll}$), energy-density profile, or uniform box—the sign, magnitude, and $p_T$ dependence of $v_1$ can vary substantially, underscoring the critical role of early-time heavy-quark positioning in shaping final-state observables.  
The magnitude of $v_1$ increases with stronger drag, reflecting its sensitivity to medium transport properties. Compared to the $p_T$-integrated $v_1$ of light-flavor hadrons, the heavy-quark signal is roughly an order of magnitude larger across the centralities studied (up to $b = 6$~fm), arising from the asymmetric drag experienced in the lopsided energy density profile. In conjunction with other observables such as elliptic flow ($v_2$) and the nuclear modification factor ($R_{AA}$), $v_1$ offers a complementary probe of heavy-quark interactions in the quark-gluon plasma.  

These findings suggest that similar effects may arise in other asymmetric systems, such as Pb+O, where geometric asymmetry could generate analogous heavy-flavor $v_1$ signals. Additionally, asymmetric collisions produce highly nonuniform electromagnetic fields. Incorporating electromagnetic effects in future studies could illuminate the interplay between medium-induced and electromagnetic contributions to heavy-quark directed flow, complementing prior investigations that focused mainly on symmetric collisions~\cite{Das:2016cwd,Jiang:2022uoe,Panda:2026kko}.  

In summary, heavy-quark $v_1$ in asymmetric collisions emerges as a sensitive probe of both the initial geometry and transport properties of the medium. Experimental measurements of heavy-flavor directed flow can directly constrain the early-time spatial distribution of heavy quarks and their interactions with the quark-gluon plasma, offering new insights into heavy-flavor dynamics in high-energy nuclear collisions.

\begin{acknowledgments}
 Ankit Kumar Panda is supported by Central China Normal University and by NSFC Grants No.12435009.  Tribhuban Parida acknowledges support from the AGH University of Krakow and the Polish National Science Centre
grant:2023/51/B/ST2/01625.
\end{acknowledgments}

\bibliographystyle{apsrev4-2}
\bibliography{ref}

\end{document}